\documentclass[preprint]{aastex}
\def\lsim{\lower.5ex\hbox{$\; \buildrel < \over \sim \;$}}
\def\gsim{\lower.5ex\hbox{$\; \buildrel > \over \sim \;$}}

\def\t{\ifmmode {\tau} \else $\tau$ \fi}

\def\ref{\noindent \hangafter=1 \hangindent=0.7 truecm}

\def\cm{\ifmmode {\rm cm}^{-1} \else cm$^{-1}$ \fi}
\def\s{\ifmmode {\rm s}^{-1} \else s$^{-1}$ \fi}
\def\cc{\ifmmode {\rm cm}^{-3} \else cm$^{-3}$ \fi}
\def\cs{\ifmmode {\rm cm}^{-2} \else cm$^{-2}$ \fi}
\def\g{\ifmmode \gamma \else $\gamma$\fi}
\def\G{\ifmmode \Gamma \else $\Gamma$\fi}

\def\kms{\ifmmode {\rm km\ s}^{-1} \else km s$^{-1}$\fi}

\begin{document}

\title{Host Galaxies, Obscuration and Nuclear Structure of Three Nearby Compact
Symmetric Objects}

\author{Eric S. Perlman \altaffilmark{1,2,3}, John T. Stocke \altaffilmark{4}, 
John Conway\altaffilmark{5}, Chris Reynolds \altaffilmark{6,7}}

\altaffiltext{1}{Current Address: Department of Physics, University of Maryland, 
Baltimore 
County, 1000 Hilltop Circle, Baltimore, MD  21250.}

\altaffiltext{2}{Department of Physics and Astronomy, Johns Hopkins
University, 3400 North Charles Street, Baltimore, MD 21218.}

\altaffiltext{3}{Space Telescope Science Institute, 3700 San Martin Dr., 
Baltimore, MD 21218.}

\altaffiltext{4}{Center for Astrophysics and Space Astronomy, University of 
Colorado, Campus Box 389, Boulder, CO 80309-0389.}

\altaffiltext{5}{Onsala Space Observatory, Chalmers Institute of Technology, 
Onsala, SWE.}

\altaffiltext{6}{Joint Institute for Laboratory Astrophysics, University of 
Colorado, Campus Box 440, Boulder, CO 80309-0440.}

\altaffiltext{7}{Hubble Fellow}

\email{perlman@jca.umbc.edu}

\newpage

\begin{abstract}

We present 3-band HST imaging of three nearby ($z\leq 0.1$) compact symmetric
objects (CSOs): 4C 31.04, 1946+708 and 1146+596.  These objects were chosen on
the basis of proximity to Earth as  well as HI 21 cm line absorption.  The
inner $H$-band isophotes of these galaxies are well fit by ``nuker'' models,
typical of nearby ellipticals.  Each shows a significant flattening in the 
isophotal profile at radii $\sim 0.5''$, as well as  significant variations in
ellipticity and PA.  However, as previous authors have noted, neither is
uncommon for elliptical galaxies.  All three objects show modest departures
from nuker law models at radii of  1-5 $h_{60}^{-1}$ kpc. Each galaxy shows
large, well distributed dust features, which are somewhat concentrated in the
nuclear regions in features which resemble disks or tori.  We find that the
amount of dust in these galaxies is about 10 times higher than normal for
ellipticals and radio galaxy hosts.  The major axes of the nuclear dust disks
tend to be oriented roughly perpendicular to the radio axis.  One galaxy, 4C
31.04, exhibits bright nuclear regions well-aligned with the radio axis, while
another, 1146+596, shows a significant near-IR excess resembling a stellar bar
along its dust disk.  The combination of outwardly normal isophotal profiles
with significant variations in PA and ellipticity is consistent with the host
galaxies being relatively recent merger remnants, with the mergers having
occurred $\gsim 10^8$ years ago.  Such a merger could have ``triggered'' the
onset of the current active phase seen in these objects, but our data require a
significant time delay between the merger event and the onset of nuclear
activity.  However, these data are also consistent with the  hypothesis that
the onset of nuclear activity in radio galaxies is due to relatively minor
``feeding'' events and/or the formation of ``bars within bars'', events which
would disturb the internal kinematics only slightly.

\end{abstract}

\newpage

\section{Introduction}

One of the most enduring mysteries concerning radio galaxies is the trigger for
the onset of the active phase.   This multifaceted issue strikes at the heart
of our understanding of the AGN phenomenon, and yet  it is difficult to address
because of the high resolutions required to peer into the innermost regions of
radio galaxies, and also because of the rarity of the youngest active galaxies.

Observational data now point to the Compact Symmetric Objects (CSOs) and
GigaHertz Peaked Spectrum (GPS) radio galaxies as the most likely candidates
for being ``young'' radio galaxies.  The distinguishing property of the CSO and
GPS classes (heretofore we will use the CSO moniker to refer to both classes;
as reviewed by O'Dea 1998, they share most observational properties) is their
small lobe separations: typically tens to hundreds of parsecs.  Two very recent
lines of evidence now point directly to ages $10^{3-4}$ years, including lobe
advance speeds $\sim 0.1-0.3 c$ (Owsianik \& Conway 1998, Taylor et al. 1998,
Owsianik et al. 1998, Conway 2000) and synchrotron aging and energy supply
arguments [Readhead et al. 1996a, b; see also Murgia et al. (1999) for  similar
analyses on CSSs (compact steep spectrum sources) and CSOs].  This ``youth''
scenario currently appears much more viable than the alternate ``frustration''
model (e.g., Fanti et al. 1990), which requires an unphysically large amount of
material ($\sim 10^{11} M_\odot$) within the inner $\sim 100$ pc (Readhead et
al. 1994, DeYoung 1993).

It is natural to ask how (or if) the host galaxies of CSOs differ from those of
large-scale, powerful radio galaxies (PRGs).  The answer to this question is
not yet apparent.  On one hand, there should be a continuum of properties, and
along these lines recent papers indicate that both CSOs and larger-scale PRGs
reside in bright elliptical galaxies with predominantly old stellar populations
(De Vries et al. 1998a, 1998b, 2000; Snellen et al. 1996, 1998b; O'Dea et al.
1996).  Yet if mergers play a significant role in triggering nuclear activity
(see, e.g., Wilson \& Colbert 1995), one might expect more dense nuclear 
interstellar media and unusual nuclear kinematics in the youngest AGN.   Along
these lines, a few CSOs appear to be highly reddened (O'Dea et al. 1996, De
Vries et al. 1998a), their narrow emission line properties argue for higher gas
pressures than large-scale PRGs (Fosbury et al. 1987), and 75\% of the HI
absorption detections among radio galaxies in the unbiased survey of van Gorkom
et al. (1989), have turned out to be CSOs (although it may simply be that CSOs
are more favorable targets for single-dish HI surveys such as that of van
Gorkom et al.; see also Conway 1996).

To date 13 CSOs have been observed with HST.  Ten are at $z > 0.1-0.8$ and have
been analyzed by McHardy et al. (1994), Perlman et al. (in prep), De Vries et
al. (2000), Evans et al. (1999) and Scoville et al. (2000). Unfortunately, in
$z>0.2$ objects the limiting resolution of HST is a few hundred to one thousand
parsecs, which is insufficient to probe structures comparable to the size of
the radio sources in CSOs. 

It was with the goal of observing nuclear structure that we observed three CSOs
at $z\leq 0.1$ with HST.  These objects were selected because of their
proximity to Earth, but also because they had been detected in HI absorption
(van Gorkom et al. 1989, Mirabel 1990, Peck, Taylor \& Conway 1998).  The
presence of HI absorption was viewed to be particularly important to test the
Wilson \& Colbert (1995) hypothesis that the ``turn-on'' of activity in
radio-loud AGN is a product of merger events between possibly active spiral
galaxies, because a powerful AGN could blow out dense nuclear gas fairly
early in its lifetime via a combination of ram pressure from the radio jet and
radiation pressure.  


The paper is laid out as follows.  In \S 2 we discuss the observations and data
reduction procedures.  In \S 3 we show the images and discuss each galaxy's
morphological and environmental characteristics, including the  location of
dust lanes and emission features, and relationship to the radio emission.   In
\S 4, we discuss the fitting of surface brightness profiles and isophotal
models.  In \S 5 we discuss the dust distribution and derive extinction maps. 
Finally, in \S 6 we discuss the impact of our results on our knowledge of CSOs
and the formation of radio galaxies.

Throughout the paper we assume $H_0 = 60 ~h_{60}~{\rm ~km ~s^{-1} ~Mpc^{-1}}$
and $q_0 = 0.1$.  Table 1 lists the corresponding physical scale, in  parsecs
per arcsecond, for each object.

\section{Observations and Data Reduction Procedures}

HST observations were done in three bands, WFPC2 F450W and F702W, and NICMOS
F160W (corresponding roughly to $B$, $R$ and $H$ bands).  The WFPC2 images were
taken with the PC1, while the NICMOS images were taken with the NIC1 chip. In
Table 1, we give a log of the observations. Due to the high absorbing columns
in these sources (plus the typically red colors of both elliptical and spiral
galaxies), we chose to integrate significantly longer in F450W than in F702W. 
The longest integrations were done in F160W, however, because of the HST's
lower sensitivity in the near-IR. The wider field of view of the WFPC2 alllowed
us to image companion galaxies in the WF chips.

The WFPC2 data were reduced using the best available dark and flat-field
images, using the IRAF/STSDAS tasks CALWP2 and CRREJ according to standard
recipes.  SYNPHOT values in the images headers were used to flux calibrate 
and transform the images to the standard $B$ and $R$ bands.

The NICMOS images were reduced using the best available dark and flat-field
images in CALNICA.  Since no dithering was done, these images are affected by
``grot'' and warm pixels.  We used  UNPEDESTAL (van der Marel 1998) to equalize
pedestal levels in the four quadrants of the NICMOS chip; however, we chose not
to subtract out the pedestal entirely because of the difficulty of measuring
the pedestal when a galaxy occupies most or all of the chip.  We resampled the
NICMOS image to square pixels (the NICMOS pixels are slightly rectangular), but
did not correct for distortion since our main interest in these observations is
the central $\sim 5''$ of each galaxy, for which the corresponding errors in
registration are $<<1$ pixel (NICMOS team 1999).  The NICMOS images were
flux-calibrated and transformed to the standard $H$ band using the SYNPHOT
information in their headers.

Once the images were reduced to this point, we registered the NICMOS and WFPC
images.  In so doing, we used $0.0455''$ pixels and assumed the position and
orientation of the nucleus on the WFPC images as fiduciary.  Two-band  (B$-$R)
and (R$-$H) color maps were then constructed from the flux-calibrated images
(using mag(Vega) = 0.0 in each band).

\section{Host Galaxy Characteristics}

In Figures 1-3, we show $R$, $(B-R)$ and $(R-H)$ images.  Smaller regions are
shown in the color index images than in the F702W image, to bring out detail. 
Two views of the F702W image are shown in each of Figures 1-3: the top left 
panel in each figure shows the entire WFPC2 ``chevron'', whereas the  top right
panel shows the inner $20''$ (for 1146+596) or $10''$ (for 4C 31.04 and
1946+708).  

Below we give a general discussion of the morphological features we find for
each object.  We also give some background from the literature on each.

{\bf 1146+596.}  This object (Figure 1) is the nearest of our targets, and is
one of the least radio-luminous CSOs.  Its milliarcsecond-scale continuum
structure is double-lobed, and has been monitored by Taylor, Wrobel \&
Vermeulen (1998), who detect component advance speeds of $0.23 \pm 0.05 c$.  
Unlike most CSOs, 1146+596 has some faint kpc scale radio emission (Wrobel,
Jones \& Shaffer 1985), which has led some authors (e.g., Conway 1996) to
speculate that it might be an example of a recurrent radio source (cf. Reynolds
\& Begelman 1997).  HI observations of this source were made with the VLBA by
Peck \& Taylor (1998).  

The host galaxy is NGC 3894, previously described as an E4 (de Vaucouleurs
1976, 1991), E (Sandage \& Tammann 1981) or S0 (Nilson 1973).   There is a
spiral companion $\sim 2'$ away (NGC 3895, $z=0.0105$, $\sim 25$ kpc  projected
distance) which is visible on the sky survey plates, and various fainter dwarf
galaxies appear on the WFPC2 mosaic (Figure 1a).  1146+596 is the least
luminous of the three hosts (Table 2), with an absolute magnitude of
approximately $L^*$. Ground-based observations found the rare combination of 
``disky'' isophotes (i.e., positive isophotal moment $a_4$), at radii $\lsim
5''$, but ``boxy'' isophotes (i.e., negative $a_4$) at radii $\gsim 10''$
(Bender, D\"obereiner \& M\"ollenhoff 1988; Nieto \& Bender 1989; Nieto,
Poulain \& Davoust 1994). Bender and collaborators also reported an anomalous
rotation curve and a M/L ratio about 3 times higher than typical for their
sample.   The work of Kim (1989) revealed a kiloparsec-scale dust disk, with
associated H$\alpha$ line emission.  The source has been detected with IRAS,
and from those data Forbes (1991) calculated a mass of $1.8 \times 10^7
M_\odot$ of cold dust, very large for an elliptical galaxy.  The object was
first detected in HI absorption by Dickey (1986), and also appears in the
sample of van Gorkom et al. (1989).

In our WFPC2 data, the nuclear emission appears rather irregular in shape, due
to the presence of two dust lanes.   These dust features are very narrow ($\sim
0.1''$), extend $\sim 300 $ pc on either side of the nucleus, and are oriented
nearly perpendicular to the arcsecond and milli-arcsecond radio axes (Figures
1c, 1d).   The dust lanes are considerably redder than the surrounding galaxy: 
$(R-H) \sim 1.3$ mag compared to $\sim 0.8$ mag, and  $(B-R) \sim 2-2.5$ mag
compared to $\sim 1.5$ mag.  The observed colors and geometry are consistent
with the  northwestern and southeastern branches representing (respectively)
the foreground and background (with respect to the nucleus) halves of a dust
torus at inclination angle $\sim 70^\circ$, an interpretation supported by the
HI distribution and kinematics (Peck \& Taylor 1998). There are also two dusty 
``loops'' on either side of (and connected to) the dust disk, as well as patchy
features in the outer parts of the galaxy. When elliptical galaxy isophotes are
subtracted from the F160W image, there is a ``bar''-like feature in the
residuals (Figure 1e) along the dust disk, suggesting either star formation
within the region of the dust disk or a non-axisymmetric potential in the
nuclear regions (which could drive a radial inflow of gas; Shlosman, Frank \&
Begelman 1989).  This ``bar'' also appears in the $(R-H)$ color map. The HI
opacity appears to be higher on the northwestern lobe (Peck \& Taylor 1998);
this is consistent with the redder colors observed in that part of the dust
disk.

{\bf 4C31.04.} This object (Figure 2) is associated with the galaxy MCG 5-4-18.
The HI absorption was first detected in single-dish observations by van Gorkom
et al.  (1989),  and Mirabel (1990) also noticed a high-velocity cloud which
later VLBA observations revealed was projected against the galaxy's center
(Conway 1999).  The source was detected with IRAS (Impey, Wynn-Williams \&
Becklin 1990), yielding evidence of considerable dust emission; those authors,
however, did not calculate a dust mass.    The VLBI radio morphology of the
object has been discussed by Conway (1996, 1999), but there is as yet no
estimate of lobe advance speeds for  this object. 

There is a companion galaxy $20''$  away (MCG 5-4-17, $z=0.055$, 20 kpc
projected distance) which is visible on both sky survey plates and our WFPC2
image (Figure 2a), as well as various fainter galaxies which could well form a
group associated with the object. Neither the host to 4C31.04 nor its spiral
companion  shows ``tidal tails'' or other large-scale evidence of interaction. 

The optical galaxy is about two magnitudes brighter than $L^*$ (Table 2).  It
is permeated with obscuration features in both the nucleus and outer regions.
The nucleus extends perpendicular to the majority of the nuclear dust, with
cone-like features very closely aligned with the radio structure.  The western
extension is somewhat redder [$(R-H) \approx 1.5$ mag] than the eastern one
[$(R-H) \approx 1$ mag].  By comparison the mean color of the galaxy is
$\langle (R-H) \rangle  \approx 1.25 \pm 0.21$ mag.    The nuclear obscuration
is concentrated in two regions.  The first is a very red [$(R-H) \approx 2$
mag] disk-like feature, oriented close to 90$^\circ$ from the radio axis, which
extends  $\sim 500$ pc to the north, and $\sim 1$ kpc to the south of the
nucleus.  There are also two 500 pc long ``arms'' which are less reddened  than
the disk.  These extend both east and west before turning to a more southerly
direction, and connect with the southern part of the disk.  A disk of similar
size and orientation to the one shown in these images, was predicted by Conway
(1999) based on VLBA HI maps.  Those maps show that the HI opacity is higher on
the Eastern side, but patchy obscuration extends to the West side. A
significant amount of patchiness  is likely required to reconcile the obscuring
features observed in the HI and HST images, although we note that the size
scales are rather different. The HI observations were modelled by a disk of
atomic gas of radius 500-1000 pc and thickness 100 pc, inclined within
$20^\circ$ of edge-on.  

{\bf 1946+708.} This object (Figure 3) is our most distant target. It was first
detected in HI by Peck, Taylor \& Conway (1998; see also Conway 1996).  Taylor
\& Vermeulen (1997) have presented an excellent discussion of multi-year VLBI
monitoring of this source, which constrains the angle between the radio jet
axis and our line of sight to $65-80^\circ$. There is a bright companion galaxy
about $1'$ away (67 kpc projected distance, redshift not listed in NED), which
is visible on the WFPC image, as well as a number of fainter companions which
may form an associated group (although the redshift of these objects is not
known).  

The host galaxy is about two magnitudes brighter than $L^*$ (Table 2).   The
nuclear regions are somewhat redder than the outer regions of the galaxy
[$(B-R) \sim 2-3$ mag compared to $\sim 1.5-2$ mag]. This galaxy exhibits
nuclear obscuration, which appears to have an inclined, disk-like morphology.
The disk is about 600 pc in radius and its major axis could be about
$60^\circ$ from the radio axis. 

The HI absorption data for this object are complicated: while the highest HI
opacities are observed NE of the nucleus, this is only because that component
is very narrow in velocity space.  Much higher column densities, in fact, are
seen against the core and SW of the nucleus (Peck, Taylor \& Conway 1999),
where the velocity width of the absorption is much higher. Taylor \& Vermeulen
(1997) and Peck et al. (1998) cite this as evidence that the NE jet is
approaching.  Free-free absorption has also been detected against the SW jet,
suggesting that some of the obscuring material is ionized.  Both our HST images
and the radio HI data can be explained  by an inclined disk, but the pattern of
obscuration is not what one would expect because the highest optical
obscurations are not seen at the same point as the highest HI columns. The
obscuring material is therefore likely to be azimuthally non-uniform, perhaps
also with a varying gas to dust ratio.  

\section{Isophotal Profiles}

We fit isophotes for the host galaxies using ELLIPSE and BMODEL in IRAF.  In
doing so, we excluded obvious isophotal features such as the aligned nuclear
emission `cones' observed in 4C31.04 or the bar observed in 1146+596. This
process was done iteratively, as some features did not immediately evidence
themselves before the initial fitting.  In Figure 4, we show the isphotal
profile for each galaxy.  This figure includes plots of surface brightness,
ellipticity, PA, and moments.  The isophotes shown in Figure 4 represent the H
band data; while isophotes were also extracted from the R and B band images,
the  H band data are less affected by dust (cf. \S 5) and therefore are more
representative of the mass distribution of the galaxies.

Figure 4 also shows several other aspects of the isophotal profile: PA and
ellipticity as well as isophotal moments.   As can be seen, all three objects
display significant changes in ellipticity and PA within the range of
semi-major axes shown.   With the possible exception of the large-scale change
in ellipticity in 4C31.04, all of the features seen in these plots can be
traced to the images shown in Figures 1-3 (recall, however, that a few obvious
features were excluded from the fits).  It is likely that some of the
largest-ellipticity isophotes extracted in 4C31.04 (i.e., at nuclear distances
$<0.4''$) are a byproduct of flagging the nuclear emission `cones', so
we exclude them from further discussion. Note, however, that this cannot
explain the larger scale (at radii up to $\sim 4''$ or 4000 pc) changes in
ellipticity observed, as the ``cones'' extend over less than $1''$. 
Importantly, the amount of variation we see in PA and ellipticity is not
unusual for elliptical galaxies, as shown by the data of Bender et al. (1988),
who give many examples of objects where (for example) isophotal twists of
several tens of degrees, or changes in ellipticity from 0.1 to 0.4, are seen.

The surface brightness profiles of these galaxies cannot be well-fit by
generalized models of the form $I(r) \propto  \exp(-\alpha r^{\beta})$.  In
this formulation, $\beta = 1$ represents an exponential disk, whereas $\beta =
0.25$ represents the classical DeVaucouleurs' profile.  All three show
essentially power law surface brightness distributions at radii larger than 1
arcsecond, with significant flattenings at small radii (respectively $\sim
0.7''$ in 1146+596, $\sim 0.5''$ in 4C31.04 and $\sim 0.3''$ in 1946+708). 
This is  a common feature among elliptical galaxies, and a signature of `nuker'
models (Byun et al. 1996, Faber et al. 1997), which represent the best fits to
the nuclear regions of nearby elliptical galaxies at HST resolutions (see also
Rest et al. 2001).  The nuker models have the mathematical form

$$I(r) = I_b ~2^{(\beta-\gamma)/\alpha}~{r_b \overwithdelims () r}^\gamma 
~\left [ 1 + {r \overwithdelims () r_b}^\alpha \right ]
^{(\gamma-\beta)/\alpha}, \eqno (1)$$

{\noindent where $r_b$ is the point of maximum curvature in log-log
coordinates, $I_b$ is  the surface brightness at $r_b$,  the asymptotic
logarithmic slope inside $r_b$  is $-\gamma$, the asymptotic outer slope is
$-\beta$, and $\alpha$ parametrizes the sharpness  of the break.} 

Nuker models were fit to the isophotal profiles using an IDL program that
employs the CURVEFIT utility, which uses a generalized, weighted least-squares
fitting algorithm.  Initial guesses for $\gamma$, $\beta$, $r_b$ and $I_b$ were
determined using by-eye estimation.   Gaussian errors were used to weight the
data.  We did not extract isophotes at semi-major axes smaller than $\sim
0.2''$ as this is approximately the FWHM of the HST PSF in H band.

The nuker model fits are overplotted on the surface brightness profiles in
Figure 4.  As can be seen, the nuker models fit well for all three objects; the
best fit parameters are given in Table 2.   All three show deviations at large
radii: in 4C 31.04, there is a clear excess at radii $\gsim 3.5''$ (3500 pc),
in 1946+708 the nuker model overpredicts the observed isophotes at radii $\gsim
1.5''$ (2500 pc), and in 1146+596 we see a slight excess above the nuker models
at $r\gsim 4''$ (900 pc).  

In Figure 5, we show the $\alpha, \beta, \gamma, r_b$ values of these objects
compared to the sample of Faber et al. (1997).  As shown, by this measure the
surface brightness profiles of these objects are not visibly abnormal compared
to the samples of Faber et al. (1997) and Rest et al. (2001).   1946+708 
appears to be what Faber et al. would call a  ``power-law'' galaxy (despite
having a significant, if small, break at $\sim 0.3"$), while 4C31.04 and
1146+596 seem to be intermediate between the ``core'' and ``power-law'' classes
of Faber et al.  However, all of these galaxies, and particularly 4C 31.04 and
1946+708, are more distant than any of the objects in the Faber et al. (1997)
sample.  As noted by Faber et al., in distant galaxies  bulge characteristics
`blend' with those of the larger galaxy.  This would cause the fitted values of
$\gamma$ and $r_b$ to be larger than they truly are (see their Figures 2 and 3,
where Faber et al. simulate the effect of fitting the profiles of M31 and M32,
placed at the distance of the Virgo cluster).  If their significantly larger
distance is accounted for, it is likely that all  three are ``core'' galaxies
similar to other bright $L>L^*$ ellipticals.

The moments of the surface brightness distribution, $a_3, ~b_3, ~a_4, ~b_4$ 
(bottom four panels in each column of Figure 4), for these objects all change 
sharply at very close to the same radii where flattenings in the isophotal 
profiles are seen (above). In all three objects, we observe the moment $a_4$ 
become sharply negative at small radii. The other moments also show significant
departures in  the inner regions of each galaxy.  1146$+$596 shows a sharp
increase in  the moments $a_3, ~b_3$, and $b_4$ at radii $\sim 0.7''$, but at
radii $\sim 0.3''$",  $b_3$ and $b_4$ decrease sharply and become  negative. 
This is consistent with the near-IR 'bar' observed in this object,  although
the variations at $0.3''-0.7''"$ may indicate that some obscuration still 
remains in the $H$ band image.  By contrast, in 4C31.04, we observe fairly 
smooth, monotonic decreases in all four moments, beginning at radii $\sim 
0.5''''$.  This pattern is indicative of the nuclear emission 'cones' observed
in  this object.  In 1946+708, we observe small increases in $a_3$ and $a_4$ at
radii  $0.4''-0.6''"$, and then decreases in all four moments at radii
$<0.4''"$.  These  features are more difficult to trace, because unlike 4C31.04
and 1146+596, there  are no obvious features in the H band residuals of this
object.  However, these  changes are somewhat similar in pattern to those
observed in 1146+596, and as can be seen in the $(B-R)$ and $(R-H)$ color
images, 1946+708 exhibits a nuclear dust  disk at these radii, as does 1146+596
at radii where sine-wave like variations  are seen in the moments.  If this
interpretation is correct, we might expect  that some obscuration remains in
these objects at $H$ band.  This interpretation is also consistent with the
small changes in ellipticity and PA observed at similar radial distances in
1946+708.

These objects also show some larger scale isophotal features, but as can be 
seen in Figure 4, the departures in the isophotal moments are much smaller than
in the nuclear regions.  At $r>1''$, 4C 31.04 appears to have a ``disky''
profile, consistent with the presence of two larger-scale, spiral shaped  dust
lanes noted earlier.  Diskiness was also previously noted in 1146+596 at radii
of $\sim 2-10''$ by Bender et al. (1988); we see this in our data as well, but
we are not sensitive to the change to boxy isophotes they noted in this object
at radii $>10''$. In both 4C 31.04 and 1146+596, we also see significant
departures in the other moments at these larger radii.  

The variations that we see in all the isophotal moments at small radii indicate
that the kinematics in the inner regions of these objects are likely to be
irregular.  This is to be expected in "boxy" (i.e., $a_4<0$) objects; as noted
by Nieto \& Bender (1989), boxiness tends to be correlated with large kinematic
anisotropies.  Interestingly, Bender \& Nieto (1989) note that boxy isophotes 
tend to predominate in active elliptical galaxies on kiloparsec scales, thus
possibly linking isophotal and kinematical anisotropies to the feeding of AGNs.
By contrast, in "disky" (i.e., $a_4>0$) galaxies, the usual kinematic signature
is a rotationally flattened disk, but much less prominent anisotropy (Bender
1988).  Thus we can feel secure in saying that by far the most prominent
kinematic feature of these objects is likely to be large anisotropies at small
radii. We will return to this and related topics in \S 6.

\section{Obscuration Features and Distribution}

We have used the multi-color HST images to obtain estimates of the total
extinction in $B$ band, $A_B$ (see Figure 6).  In so doing, we made three
assumptions. The first assumption is that the average colors of each object are
taken as representative of the unextincted stellar population of the host. 
Since none of these objects appear to be disk-dominated objects seen along
their disks, this should be very nearly true if these galaxies follow an
extinction law similar to that for our own Galaxy (see Mathis 1990), where by
$H$-band the extinction is only 6\% of what it is in $B$-band.  We also assume
a covering factor of unity, the most conservative  assumption, but it is likely
that there is some variation in the covering  factor, just as in our own
Galaxy.  Our final assumption is that the overall galaxy colors are dominated
by continuum emission, rather than lines.  Without emission line imaging we
cannot comment on the validity of this assumption.   Once these assumptions are
made, we can then compute $A_B$ values for each pixel $(i,j)$  via 

$$(A_B)_{ij} = (B-H)_{ij} - \langle (B-H) \rangle. \eqno (2) $$


The nuclear dust disks represent the highest-extinction features seen in each
host galaxy.  In 1146+596, the northwestern branch of the dust disk reaches
$A_B = 0.9-1.1$ mag, while the southeastern branch reaches somewhat lower
values ($A_B = 0.7-0.9$ mag).   There is considerable extinction as well in
between the two visible branches of the central dusty region ($A_B = 0.5-0.7$
mag), suggesting that the nuclear dust torus is filled, rather than open.  
Somewhat larger extinction values are seen in the disks of 4C31.04 and
1946+708.  In 4C 31.04, the disk appers nearly edge-on, with  $A_B = 1.5-2.3$
mag, peaking southwest of the galaxy's  nucleus.  In  1946+708, which appears
to have a somewhat inclined disk, we see $A_B$ peaking at about 2.0 mag at the
northern extremum of the disk, and a much larger variation in extinction,
varying between 0.4 and 1.8 magnitudes, with the minimum value of $A_B$ at the
disk's southwestern edge.   

We find lower $A_B$ values in the dust ``loops'' of 1146+596 and  4C
31.04, than in their respective nuclear dust disks.  In 4C 31.04, the 
extinction values in the ``loops'' range from $A_B = 1.0-1.3$ mag in two spots,
to the northwest of the nucleus and where the loop intersects with the northern
extension of the disk, to values as low as $0.1-0.2$ mag east and southeast of
the nucleus.  By comparison, in 1146+596, the extinction in the ``loops''
ranges from $A_B = 0.4-0.5$ mag in the loop to the north  of the nucleus to
only $0.15-0.3$ mag south and east of the nucleus.

1146+596 and 4C 31.04 also exhibit extinction features in their
outer regions.  In 1146+596, these are patchy, and have moderate $A_B$ values
($A_B = 0.3-0.7$ mag). In 4C31.04, the large-scale extinction features are
distributed in an S-shaped lane, extending outwards from the nuclear disk
on both sides for up to 4 kpc.  The values of $A_B$ in this feature range from
0.7--1 mag in the northern lane to 0.2--0.5 mag in the southeastern lane.

The $A_B$ values can be transformed into HI columns by assuming the extinction
law of Mathis (1990), and the $E(B-V) \rightarrow N(H)$ relationship of Bohlin
et al. (1978), viz. $N_H = 5.8 \times 10^{21} {\rm ~cm^{-2} ~mag^{-1}}$.  
Total dust masses were obtained by integrating the  $N(H)$ maps and a
dust-to-gas ratio similar to that of our own Galaxy ($\sim 100$ by weight).
This yields the values given in Table 3.  As can be seen, all three of these
galaxies have significant HI columns and dust masses typical of spiral
galaxies.  By comparison, ellipticals typically have dust masses an order of
magnitude lower (Goudfrooij et al. 1994a, b), although there are notable
exceptions, including some large scale FR II radio sources (de Koff et al.
2000; see also \S 6).  We note in the captions to Figure 6 what the $A_B$
greyscales of each extinction map corresponds to in terms of $N_H$, under the
above assumptions.

The $N_H$ maps allow us to estimate the $T_s/f_c$ values on each side of the
radio source by comparing the values we calculate with $N_H$ values derived
from VLBI observations (Peck \& Taylor 1998, Conway 1999, Peck et al. 1999).  
To do this, of course, one must assume that the same gas serves as the
absorbing column for both the radio and optical emission, a nontrivial
assumption given that there is a resolution mismatch of approximately a factor
5-10 between the VLBA observation and our HST observations.  So these are
fairly rough estimates and at most we can differentiate between the two sides
of the radio source  and not on a point by point basis. Nevertheless, the
results are given in Table 3.  For all three objects, values of  $T_s/f_c \sim
100-200 $ K  are obtained, even though the dust and gas are within $\sim 100$
pc of the nucleus.  For two objects, 1146+596 and 4C 31.04, our results are
roughly consistent with the assumptions of $\sim 100$ K in the literature (Peck
\& Taylor 1998, Conway 1999).  However, for  1946+708, our results are
inconsistent with the assumption by Peck et al. (1998) of $T_s \sim 8000$ K,
unless a very low filling factor is invoked. This is a significant result, even
given the resolution mismatch and different methods used in the calculation. 
To reconcile the two temperature results in a covering factor $f_c \sim 0.025$,
which we believe is unlikely given the rather uniform appearance of the
obscuring screen shown in Figure 6c.  As one can see by comparing our Figure 6c
to the maps in Peck et al. (1999) the ratio between the thickness of the
densest part of the obscuring disk and the the width of the 1.3 GHz continuum
emitting region is about 5:1, so that a rather unlikely combination of
orientations for the radio minilobes and dust clouds would be required. 
Further, we note that at $T=8000$ K and the  densities calculated for this
temperature by Peck et al. (1999), one would not expect 21 cm HI absorption due
to collisional quenching.

\section{Discussion}

The hosts of these three CSOs appear typical of nearby elliptical galaxies in
many respects. All three are ellipticals, well-fit by ``nuker'' models (Faber
et al. 1997) within $\sim 3$ kpc of the nucleus.  All three show significant
breaks to flatter cores in their surface brightness distributions, although
none actually fall into the ``core'' region as defined by Faber et al. (1997).
Instead, two objects, 1146+596 and 4C31.04, fall into the intermediate region
of the $(r_b, \gamma)$ plane, while 1946+708 falls into the ``power-law''
section of that diagram.    As noted in \S 4, however, the true values of $r_b$
and $\gamma$ for  4C 31.04 and 1946+708 are likely smaller, given their 
distance and the fact that HST barely resolves the cores in these objects.  
The hosts of 4C 31.04 and 1946+708 are about 2 magnitudes  brighter than $L^*$,
which is fairly typical of radio galaxy hosts (e.g., Martel et al. 1999, Ledlow
\& Owen 1995, Smith \& Heckman 1989, Owen \& Laing 1989), but 1146+596 is
somewhat underluminous compared to most radio galaxy hosts, with $L \approx
L^*$.

As noted in \S 4, these objects have significantly non-relaxed isophotes at 
radii of a few hundred parsecs,   with significant variations in ellipticity
and PA, as well as significant deviations in the isophotal moments.  Taken
together, this is a fairly sensitive indication that the inner regions of the
underlying galaxies are not completely relaxed (e.g., Mihos et al. 1995, Mihos
\& Hernquist 1996), even if they show no outward indications of a recent
merger.  Indeed, if the kinematic and morphological properties of elliptical 
galaxies follow the same relationship on scales of tens to hundreds of parsecs
as they do on scales of kiloparsecs, this is a likely indication of large 
nuclear kinematical anisotropies in these objects. In order to understand the
overall impact of this observation, we must assess not only how unusual such
variations are, but also assess any resolution related and other biases.  Some
caution is warranted at the smallest radii (Rest et al. 2001), i.e., at
resolutions $<0.2''$   HST's sensitivity to detecting real isophotal twists 
decreases significantly, such that for a cuspy model with constant ellipticity
and PA, ELLIPSE still would erroneously detect some changes in ellipticity and
PA on scales smaller than $0.2''$.  This concern is particularly applicable to
1946+708, which has an a large change in ellipticity at $0.3''$.  Further, we
must also exclude from our analysis the isophotes within $\sim 0.4''$ for 4C
31.04 because they are heavily affected by the masking of the nuclear emission
``cones'' in that object (cf. \S 4)

With these caveats in mind, then, we are left with 2 objects (4C31.04 and
1146+596) which have ellipticities that vary between 0.10 and 0.40 in the inner
2-4 kpc, and one (1946+708) which shows much smaller variations. Similarly,
4C31.04 and 1146+596 exhibit major isophotal twists (by several tens of
degrees) while 1946+708 exhibits much less variation in PA.  Such variations in
ellipticity and PA are not at all uncommon in the nearest elliptical galaxies
on the 100-1000 pc level (Bender et al. 1988). In 1146+596, we see an
ellipticity that decreases fairly steadily in the inner few hundred parsecs. 
This is consistent with the ``diskiness'' found on $2-10''$ scales by Bender et
al. (1988) in ground-based data. The pattern of the variations seen in 4C31.04
is somewhat reversed; however, this could  still be consistent with significant
``diskiness'' on 500-2000 pc scales (as indicated by the positive values of
$a_4$ we observe at these scales) in a more face-on orientation, and it would
be consistent with the larger scale dust lanes in this object noted in \S\S 3
and 5.  

These galaxies are somewhat unusual compared to the hosts of most radio
galaxies in the amount of nuclear and galaxy wide dust.  Martel et al. (1999)
analyzed HST WFPC images of 46 $z>0.1$ 3CR radio galaxies, and further analyses
of the dust properties of the sample were done by de Koff et al.  (2000) and
Martel et al. (2000).    About half of the 3CR galaxies have some sort of dust
feature, including irregular lanes, filaments, nuclear or large-scale
disks above a threshold detection level of $\sim 3 \times 10^4 M_\odot$ in
dust.   That amount of dust is consistent with the amounts of dust found in
non-active elliptical galaxies by Goudfrooij et al. (1994a, b) and Tran et al.
(2001).  As can be seen, therefore, these objects have dust masses $0.5-1$ {\it
dex} higher than typical for elliptical galaxies and radio galaxy hosts,
although 9/46 3CR hosts are comparable to 1146+596, 4C 31.04 and 1946+708. 
Interestingly, it is not just the smallest 3CR sources that have substantial
dust masses: e.g., 3C 46, 3C 236 (but see below) and  3C 306.1 are all
large ($\gsim 1 ~h_{60}^{-1}$ Mpc) and have $M_{\rm dust}> 10^5 M_\odot$.   The
3CR objects with prominent dust lanes also show a significant, but not
universal, anti-correlation of dust lane position angle with radio axis,
similar to these objects.  Somewhat similar results were seen in the HST
snapshot  survey of BL Lacs undertaken by Urry et al. (2000) and Scarpa et al.
(2000).  Of the 58 objects in that sample with resolved hosts, 12 were at
$z<0.1$, and only one of those (1ES1959+650) has an obvious dust feature:  a
kpc-scale disk with an estimated dust mass of $\sim 5 \times 10^5 M_\odot$, in
the same range as these CSOs (Scarpa et al. 2000).

Given the large amount of nuclear dust shown by these observations, it is
natural to wonder whether these objects are typical of CSOs in general.  While
these objects were chosen partly because of their HI absorption, we note that
this is a property which is common among CSOs, since up to 50\% of CSOs now
appear to be HI absorbers (Conway 1996, 2001).  Thus, any such bias in this
regard  is small, although observations of non-HI absorbers might be helpful to
elucidate this further. 

None of our objects show obvious nuclear point sources, consistent with their
strong nuclear obscuration.  Martel et al. (1999) found nuclear point sources
in 43-54\% of $z<0.1$ 3CR radio galaxies, and some correlation between the lack
of a point source and nuclear absorption.  Chiaberge, Capetti \& Celotti
(1999), working with a very similar sample, found a strong correlation between
the optical flux of the central compact core and the radio core flux, and an
anti-correlation between optical core flux and dust content.  

Only one other $z<0.2$ CSO has been observed with HST:  PKS 1345+12 ($z=0.122$;
Evans et al. 1999, Scoville et al. 2000, Surace et al. 1998).  Like our
targets, PKS 1345+12 was detected in HI by van Gorkom et al. (1989).  However,
it seems a much more extreme object, having also been singled out as an
ultra-luminous infrared galaxy (ULIRG) by Sanders et al. (1988), with an IRAS
luminosity of $10^{12.44} L_\odot$. The host galaxy of PKS 1345+12 has an
irregular optical morphology, with  two nuclei separated by $3.0''$ and 
prominent tidal tails; thus it is almost certainly a merger event in progress. 
In  addition, in PKS1345+12 we also see large amounts  of nuclear obscuration,
and CO (Evans  et al. 1999) and OH megamaser  (Baan, Salzer \& Lewinter 1998)
emission centered on the western nucleus, which is also the location of the
AGN. Evans et al. (1999) attempt to link the molecular gas with the feeding and
development of the AGN.  Hurt et al. (1999) reported a UV polarization of $16.4
\pm 2.6 \%$ in HST/FOC observations, with a polarization PA nearly
perpendicular to the radio axis, consistent with scattered AGN light. However
there is inadequate astrometric information to associate the polarized UV
source with the AGN.  The X-ray data are consistent with a hard power-law
spectrum plus a large ($>10^{22} ~{\rm cm^{-2}}$) absorbing column (O'Dea et
al. 1996b, Imanishi \& Ueno 1999)

Nine $z>0.2$ CSOs have been observed with HST (DeVries et al. 2000, Perlman et
al. 2001, McHardy et al. 1994). All of the DeVries et al. objects are well fit
by elliptical models, and none show  irregular isophotes (although n.b., De
Vries et al. do not discuss variations in PA or ellipticity) or strong
obscuration features, although three are significantly reddened: 0404$+$768
$(z=0.599)$, 0428$+$205 $(z=0.219)$ and 0500$+$019 $(z=0.583)$, all of which
also have close companions. But due to their much greater distances, the lack
of (e.g.) visible dust features and/or aligned structures in the remaining
sources cannot be taken as a statement that they do not exist.  In fact, only
for the very nearest of the three reddened sources in the DeVries et al. sample
(0428$+$205) could HST resolve features similar in size to those observed in
the nuclear regions of these three objects. Moreover, the DeVries et al.
observations were carried out in redder bands (NICMOS F110W and F205W) so that
their sensitivity to modest amounts of dust was much less.  PKS 1413+135,
analyzed by Perlman et al. (in prep.) and McHardy et al. (1994), appears to be
a very different object, having a clearly spiral host, but strong nuclear and
galaxy wide obscuration, molecular absorption (Wiklind \& Combes 1994, 1997),
and a hard, obscured X-ray source (Sugiho et al. 1999); the latter properties
being similar to PKS 1345+12.

In summary, the three CSOs imaged here appear to be relatively normal,  albeit
relatively dusty luminous elliptical galaxies.  They also appear to be
representative of the CSO class, although a few CSOs (e.g., PKS 1345+12 and 
PKS 1413+135) appear to be more obvious candidates for very recent merger
events.

Since mergers tend to force gas and dust towards the central regions (Mihos \&
Hernquist 1996), the large dust masses of 1146+596, 4C 31.04 and 1946+708, are 
interesting in the light of the scenario proposed by Wilson \& Colbert (1995),
whereby  the ``turn-on'' of nuclear activity is triggered by a merger event.
Also interesting is the presence of companions within 20-70 kpc projected
distance, for all three objects.   But perhaps most interesting of  all in this
light is that these objects appear to have outwardly normal isophotal profiles
(fitting nuker models rather well), but exhibit significant anomalies in PA and
ellipticity, particularly in their inner regions.  This is consistent with the
finding of Mihos (1995) that $\gsim 10^8$ years after a major merger of two
disk galaxies, the inner 3-5 kpc region of the remaining galaxy appears quite
similar to a normal elliptical in its surface brightness profile but with
remnant ellipticity changes and isophotal twists.  Thus if a major merger
triggered the activity seen in these objects, it must have occurred $\gsim
10^8$ years earlier (cf. Mihos 1995, Hernquist \& Mihos 1995, Mihos \&
Hernquist 1996).

A timescale of $>10^8$ years after a major merger is similar to the timescale
over which a $10^8 M_\odot$ black hole could double its mass while accreting at
the Eddington rate (the Eddington rate is $\sim 3 ~M_\odot~{\rm yr^{-1}}$ for a
$10^8 M_\odot$ black hole, and scales linearly with the mass), and is also of
the same order as the timescale required to accrete material into the inner few
parsecs of a galaxy ( i.e., the dynamical time of the ``nested bars''
hypothesis of Shlosman et al. 1989).  These timescales are relevant if indeed
the ``activation'' of a radio-loud AGN is directly related to spinning up of
the nuclear black hole (see, e.g., Meier 1999, Reynolds et al. 1999); a
significant increase in the black hole's angular momentum would require the
accretion of a mass comparable to the mass of the black hole.  This long time
delay between the merger and the onset of nuclear activity could also be due to
the long time required for a bound black hole binary to coalesce (or for a hard
black hole binary to form - see Begelman et al. 1980), although recent
simulations indicate that the coalescence timescale for black hole binaries is
almost certainly much longer than $10^8$ years (Merritt, Cruz \& Milosavljevic
2000; Poon \& Merritt 2001), as required by the Wilson \& Colbert (1995)
scenario. In either case, our observations require that if a major merger is
responsible for the onset of nuclear activity, a mechanism with timescale $\sim
10^8$ years must delay the production of the radio source.  Simulations of
mergers (Mihos \& Hernquist  1996) suggest that gas inflow into the nuclear
regions of the product galaxy would peak $\sim 5-10 \times 10^7$ years after
the beginning of the merger,  and continue for several $\times 10^7$ years.

A timescale of $\sim 10^8$ years is also important to our understanding of
these objects for another reason: by the time that long of a time period has
elapsed, any nuclear starburst associated with the merger will have subsided,
although some evidence may remain in the colors. We do not see any evidence of
nuclear starbursts in these objects,  with the possible exception of the
near-IR ``bar'' in 1146+596.  De Vries et al. (1998a, 1998b, 2000) also found
that the colors of CSO host galaxies at higher redshifts were consistent with
predominantly old stellar populations.  


Our findings are also consistent with more quiescent feeding mechanisms for a
pre-existing black hole.  For example,  a more distant tidal encounter could
alter the gravitational potential of the inner few kpc of these galaxies,
driving gas radially inward as in the ``nested bars'' hypothesis of Shlosman et
al. (1989).  Our discovery of a stellar bar in the nearest of these objects
is surprising and constitutes strong support for the ``nested bars'' 
hypothesis.  The presence of more distant ($20-70 ~h_{60}^{-1}$ kpc) companions
in all three cases is also important in this regard.  Tidal encounters with
these companions could also produce the isophotal anomalies we observe, and
would also be consistent with a less-violent mode for nuclear star formation,
thus lessening the effect of a nuclear superwind on the fuelling of the nascent
central engine.   If indeed tidal encounters or other more quiescent methods
are the preferred way to trigger nuclear activity in AGN, these objects may not
be in their first  active phase. In this case, we might expect to find evidence
for previous nuclear activity due to previous tidal encounters with companion
galaxies.  By contrast, in the merger scenario, relic haloes are much harder to
explain and would not be expected.  

Previous authors (e.g., Conway 1996 and sources therein) have cited the
kpc-scale radio halo of 1146+596 as evidence of a possible previous active
stage, perhaps indicative of a recurrent radio source (e.g., Reynolds et al.
1997).  Similar statements have been made for 3C 236 (O'Dea et al. 2001,
Schilizzi et al. 2001), a CSS with both compact, 2 kpc-size inner radio lobes
and a Mpc-scale radio halo.  The size of the variations in isophotal PA and
ellipticity seen in 3C 236 are similar to those we see in 1146+596, as are the
mass and relative orientation of its nuclear dust disk, and the amount of
larger-scale dust features.  3C 236 also has evidence for recent star formation
of varying ages in these dust features.  O'Dea et al. (2001) suggest on the
basis of these features that the fuel supply to the AGN in 3C 236 was
interrupted for $\sim 10^7$ years and is now restored. They argue in favor of a
non-steady transport of gas in the disk and more quiescent triggering of active
stages based on gas-dynamical processes.

In the light of these findings, further observations  of nearby CSOs are
warranted.   The results of a sensitive 90 cm VLA search for very steep
spectrum, aged radio remnants around CSOs would be an important discriminant
between the merger  and more quiescent triggering scenarios. Emission-line
imaging and spectra of these objects with HST would allow us to disentangle
their nuclear kinematics and dynamics, enabling us to relate the dust, gas
disks and any nuclear star formation to the  feeding and triggering of the
nuclear activity.  Finally, observations of many more nearby CSOs (including
objects {\it not} selected for large obscuring columns) are needed to
substantiate the results from this very small sample, and hopefully give us
more  concrete evidence  regarding the Wilson \& Colbert (1995) scenario that
radio galaxy activity is triggered by major merger events.

\begin {acknowledgments}

We wish to thank Al Schultz, Howard Bushouse and Eddie Bergeron for
considerable help with the HST data reductions.  We also wish to thank an
anonymous referee for many suggestions which significantly improved this paper.
We thank Peter Barthel, Eleni Chatzichristou, Tim Heckman, Susan Neff,  Chris
O'Dea, Alison Peck, Eric Smith and Greg Taylor for interesting discussions.
Support for proposal GO-7344 was provided by NASA through a grant from the
Space Telescope Science Institute, which is operated by the Association of
Universities for Research in Astronomy, Inc., under NASA contract NAS5-26555.
ESP acknowledges support at Johns Hopkins University and the University of 
Maryland from NASA LTSA grant NAG5-9534/NAG5-9997 and HST grant GO-8060.  CR
acknowledges support at the University of Colorado from Hubble Fellowship
HF-01113.01-98A. 

\end {acknowledgments}

\vfill\eject

\begin {references}

Baan, W. A., Salzer, J. J., \& Lewinter, R. D., 1998, ApJ, 509, 633.

Begelman, M. C., Blandford, R. D., \& Rees, M. J., 1980, Nature, 287, 307.

Bender, R., 1988, A \& A, 193, L7

Bender, R., D\"obereiner, S., \& M\"ollenhoff, C., 1988, A \& A S, 74, 385.

Bohlin, R. C., Savage, B. D., \& Drake, J. F., 1978, ApJ, 224, 132.

Byun, Y.--I., Grillmair, C. J., Faber, S. M., Ajhar, E. A., Dressler, A., 
Kormendy, J., Lauer, T. R., Richstone, D., Tremaine, S., 1996, AJ, 111, 1889.

Carilli, C. L., Perlman, E. S., \& Stocke, J. T., 1992, ApJL, 400, L13.

Chiaberge, M., Capetti, A., \& Celotti, A., 1999, A \& A, 349, 77.

Conway, J. E. 1996, The Second Workshop on Gigahertz Peaked Spectrum and
Compact Steep Spectrum Radio Sources, ed. I. A. G. Snellen, R. T.  Schilizzi,
H. J. A. R\"ottgering, and M. N. Bremer, p. 198.

Conway, J. E., 1999, NAR, 43, 509.

Conway, J. E., 2000, in {\it Lifecycles of Radio Galaxies},
ed. J. Biretta, A.  Koekemoer, C. O'Dea \& E. Perlman, NAR, in press.

Conway, J. E., 2001, IAU Symposium 205, 'Galaxies and their contents
at the highest resolution', in press. 


De Koff, S., et al. 2000, ApJS, 129, 33.

De Vaucouleurs, G., 1976, ``Second Reference Catalogue of Bright Galaxies''
(Austin:  University of Texas Press).

De Vaucouleurs, G., 1991, ``Third Reference Catalogue of Bright Galaxies''
(New York:  Springer-Verlag).

De Vries, W.,  et al. 1998a, ApJ, 503, 138.

De Vries, W., et al. 1998b, ApJ, 503, 156.

De Vries, W., O'Dea, C.,  Barthel, P., Fanti, C., Fanti, L., \&  Lehnert, M.
D., 2000, AJ, 120, 2300.

Dickey, J., 1986, ApJ, 300, 190.

Evans, A. S., Kim, D. C., Mazzarella, J. M., Scoville, N. Z., \& Sanders, D. B.,
1999, ApJ, 521, L107.

Faber, S. M., et al. 1997, AJ 114, 1771.

Fanti, R., Fanti, C., Schilizzi, R., Spencer, R., Redong, N., Parma, P., Van
Breugel, W., \& Venturi, T., 1990, A\& A, 231, 333.

Forbes, D. A., 1991, MNRAS, 249, 779.

Fosbury, R. A. L., Bird, M. C., Nicholson, W., \& Wall, J. V., 1987, MNRAS,
225, 761.

van Gorkom, J. H., Knapp, G. R., Ekers, R. D., Ekers, D. D., Laing, R. A.,
\& Polk, K. S., 1989, AJ, 97, 708.

Goudfrooij, P., De Jong, T., Hansen, L., Norgaard-Nielsen, H. U., 1994a, 
MNRAS, 271, 833.

Goudfrooij, P.,  Hansen, L., Jorgensen, H. E., \& Norgaard-Nielsen, H. U.,
1994b, A\& A S, 105, 341.

Hernquist, L., \& Mihos, J. C., 1995, ApJ 448, 41.

Hurt, T., Antonucci, R., Cohen, R., Kinney, A., \& Krolik, J., 1999, 
ApJ, 514, 579.

Imanishi, M., \& Ueno, S., 1999, ApJ, 527, 709.

Impey, C. D., Wynn-Williams, C. G., Becklin, E. E., 1990, ApJ, 356, 621.

Kim, D.--W., 1989, ApJ, 346, 653. 

Ledlow, M. J., \& Owen, F. N., 1995, AJ, 110, 1959.



Lin, H., Kirshner, R. P., Shectman, S. A., Landy, S. D., Oemler, A., Tucker,
D. L., \& Schechter, P. L., 1996, ApJ, 464, 60.

Martel, A. R., et al., 1999, ApJ, 122, 81.

Martel, A. R., et al., 2000, ApJ, in press.

Mathis, J. S., 1990, ARAA, 28, 37.

McHardy, I. M., Merrifield, M. R., Abraham, R. G., \& Crawford, C. S.,
1994, MNRAS, 268, 681.

Meier, D. L., 1999 ApJ, 522, 753.

Merritt, D., Cruz, F., Milosavljevic, M., 2000, in ``Dynamics of Star
Clusters'', ed. S. Deiters et al., in press, astro-ph/0008497.

Mihos, J. C., 1995, ApJL, 438, L75.

Mihos, J. C., \& Hernquist, L., 1996, ApJ, 464, 641.

Mihos, J. C., Walker, I. R., Hernquist, L., Mendes de Oliviera, C., \& Bolte,
M., 1995, ApJ 447, L87.

Murgia, M., Fanti, C., Fanti, R., Gregorini, L., Klein, U., Mack, K. -- H.,
\& Vigotti, M.,  1999, A \& A, 345, 769.

NICMOS team, ``NICMOS Photometry Update,'' 1999, 
http://www.stsci.edu/cgi-bin/nicmos

Nieto, J. L., \& Bender, R., 1989, A \& A, 215, 266.

Nieto, J. L.,  Poulain, P., \& Davoust, E., 1994, A \& A, 283, 1.

Nilson, P., 1973, ``Uppsala General Catalogue of Galaxies,'' Nova Acta Regiae
Soc. Sci. Upsaliensis, Ser. V:A, 1.

O'Dea, C. P., 1998, PASP, 110, 493.

O'Dea, C. P., Koekemoer, A. M., Baum, S. A., Sparks, W. B., Martel, A. R., 
Allen, M. G., Macchetto, F. D., \& Miley, G. K., 2001, AJ, in press
(astro-ph/0101441).

O'Dea, C. P., Stanghellini, C., Baum, S. A., \& Charlot, S., 1996a, ApJ, 470, 
806.

O'Dea, C. P., Worrall, D. M., Baum, S. A., \& Stanghellini, 1996b, AJ, 111, 92.

Owen, F. N., \& Laing, R. A., 1989, MNRAS, 238, 357.

Owsianik, I., \& Conway, J. E., 1998, A \& A, 337, 69.

Owsianik, I., Conway, J. E., \& Polatidis, A. G., 1998, A \& A, 336, 37.

Peck, A. B., \& Taylor, G. B., 1998, ApJ, 502, L23.

Peck, A. B., Taylor, G. B., \& Conway, J. E., 1999, ApJ, 521, 103.

Perlman, E. S., Carilli, C. L., Stocke, J. T., \& Conway, J. E., 1996,
AJ, 111, 1839.

Perlman, E. S., Stocke, J. T., Shaffer, D. B., Carilli, C. L., \& Ma, C.,
1994, ApJL, 424, L69.

Perlman, E. S., Stocke, J. T., Carilli, C. L., \& Conway, J. E., Wang, Q. D., 
Tashiro, M., \& Sugiho, M., 2001, in preparation

Poon, M. Y., Merritt, D., 2001, ApJ, 549, 192.

Readhead, A. C. S., Xu, W., Pearson, T. J., Wilkinson, P. N., \& Polatidis,
A. G., 1994,  in Compact Extragalactic Radio Sources (1994 NRAO Symposium), 
eds. Zensus, A., \& Kellerman, K., (ASP:  Provo), pp. 17-22.

Readhead, A. C. S., Taylor, G. B., Xu, W., Pearson, T. J., Wilkinson, P. N.,
\& Polatidis, A. G., 1996a, ApJ, 460, 612.

Readhead, A. C. S., Taylor, G. B., Pearson, T. J., \& Wiklinson, P. N., 
1996b, ApJ, 460, 634.

Rest, A., van den Boesch, F. C., Jaffe, W., Tran, H., Tsvetanov, Z., Ford, H.
C., Davies, J., \& Schafer, J., 2001, AJ, in press, astro-ph/0102286.

Reynolds, C. S. \& Begelman, M. C., 1997, ApJ, 488, 109.

Reynolds, C. S., Young, A. J., Begelman, M. C., \& Fabian, A. C., 
1999, ApJ, 514, 164.

Sandage, A., \& Tamman, G. A., 1981, ``A Revised Shapley-Ames Catalog of Bright
Galaxies'' (Washington:  Carnegie Institution).

Sanders, D. B., Soifer, B. T., Elias, J. H., Neugebauer, G., \& Matthews, K.
1988, ApJ, 328, L35.

Scarpa, R., Urry, C. M., Falomo, R., Pesce, J. E., Webster, R., O'Dowd, M.,
\& Treves, A., 1999, ApJ, 521, 134.

Scarpa, R., Urry, C. M., Falomo, R., Pesce, J. E., Treves, A., 2000, ApJ,
532, 740.

Schilizzi, R. T., Tian, W. W., Conway, J. E., Nan, R., Miley, G. K., Barthel,
P. D., Normandeau, M., Dallacasa, D., and Gurvits, L. I., 2001, A \& A, 368,
398/.

Scoville, N. Z., Evans, A. S., Thompson, R., Rieke, M., Hines, D. C.,
Low, F. J., Dinshaw, N., Surace, J. A., \& Armus, L., 2000, AJ, 119, 991.

Shlosman, I., Frank, J., \& Begelman, M. C., 1989, Nature, 338, 45.

Smith, E. P., \& Heckman, T. M., 1989, ApJS 69, 365.

Snellen, I. A. G., Bremer, M. N., Schilizzi, R. T., Miley, G. K. \&
Van Oje, R., 1996, MNRAS, 279, 1294.

Snellen, I. A. G., et al. 1998, MNRAS, 301, 985.

Stocke, J. T., Wurtz, R., Wang, Q., Elston, R., \& Jannuzi, B., 1992, ApJL,
400, L17.

Sugiho, M., Tashiro, M., Perlman, E. S., Makashima, K., Stocke, J. T., 
Wang, Q. D., \& Madejski, G., 1999, Astronomische Nachrichten, 320, 316.

Surace, J. A., Sanders, D. B., Vacca, W. D., Veilleux, S., \& Mazzarella, J. M.,
1998, ApJ, 492, 116.

Taylor, G. B., Readhead, A. C. S., \& Pearson, T. J., 1996, ApJ, 463, 95.

Taylor, G. B., \& Vermeulen, R. C., 1997, ApJ, 485, L9.

Taylor, G. B., Wrobel, J. M., \& Vermeulen, R. C., 1998, ApJ, 498, 619.

Tran, H. D., Tsvetanov, Z., Ford, H. C., Davies, J., Jaffe, W., van den Bosch,
F. C., \& Rest, A., 2001, AJ, in press, astro-ph/0102292.

Urry, C. M., Scarpa, R., O'Dowd, M., Falomo, R., Pesce, J. E., \& Treves, A.,
2000, ApJ, 532, 816.

van der Marel, R., 1998,  http://sol.stsci.edu/marel/software/pedestal.html

Wiklind, T., \& Combes, F., 1994, A\& A, 286, L9.

Wiklind, T., \& Combes, F., 1997, A \& A, 328, 48.

Wilson, A. S., \& Colbert, E. J. M., 1995, ApJ, 438, 62.

\end{references}

\begin{figure}

\centerline{\includegraphics*[scale=1.00]{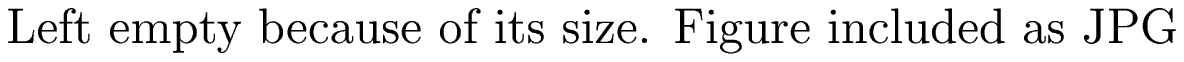}}

\caption[]{HST images of 1146+596 (also NGC3894, $z=0.0107$).  At top left, the
entire F702W mosaic  (greyscale runs from 0 to 100 ADU/pix; image is $160''$
square); top right, a blowup of the the F702W image (scale runs from 21.5
(black) to 27 mag(white)), at bottom left, the $(B-R)$ image (greyscale runs
from 1 (white) to 2.5 mag (black)), at bottom middle, the $(R-H)$ image
(greyscale runs from 0.7 (white) to 1.3 mag (black) ), and at bottom right, the
residuals from the subtraction of the azimuthally-averaged surface profile of
the galaxy in the $H$ band image (greyscale runs from 0 (white) to 0.4
ADU/pix/s (black) ). The faint ring in Figure 1e (bottom right) is an artifact
of the galaxy subtraction process, perhaps linked to residual obscuration in 
the dusty ``loops'' seen at similar radii.  North is at top and East is at left
in all panels.}



\end{figure}

\vfill\eject

\begin{figure}

\centerline{\includegraphics*[scale=1.00]{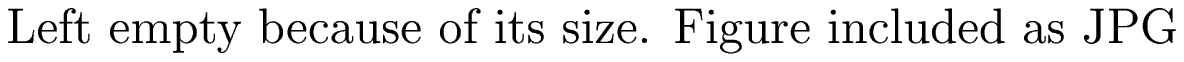}}

\caption[]{HST images of 4C31.04 ($z=0.057$).  At top left, the entire F702W
mosaic; at top right (greyscale runs from 0 to 100 ADU/pix; image is $160''$
square), a blowup of the F702W image (scale runs from 23 (black) to 28 mag/pix
(white)), at bottom left, the $(B-R)$ image  (greyscale runs from 1.5 (white)
to 3 mag (black)), and at bottom right, the $(R-H)$ image (greyscale runs from
0.7 (white) to 2.2 mag (black)). The vertical lines in Figure 2d (bottom left)
are artifacts due to  quadrant effects.  North is at top and East is at left in
all panels.}

%

\end{figure}

\vfill\eject

\begin{figure}
 
\centerline{\includegraphics*[scale=1.00]{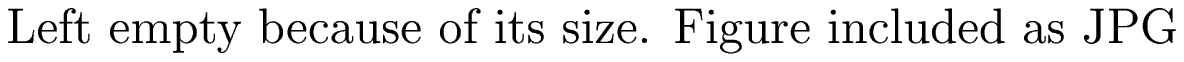}} 
 
\caption[]{HST images of 1946+708 ($z=0.101$).  At top left, the entire F702W 
mosaic (greyscale runs from 0 to 100 ADU/pix; image is $160''$ square); at top
right, a blowup of the F702W image (scale runs from 23 (black) to 28 mag/pix
(white)), at bottom left, the $(B-R)$ image (greyscale runs from 1 (white) to 3
mag(black)), and at bottom right, the $(R-H)$ image (greyscale runs from 1
(white) to 2 mag (black)).  North is at top and East is at left in all panels.} 

 

\end{figure}

\vfill\eject

\begin{figure}

\centerline{\includegraphics*[scale=0.78]{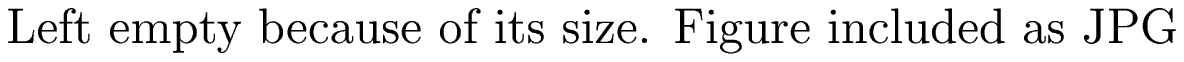}}   

\caption[]{Host Galaxy surface brightness profiles of all three objects.  At
left, 1146+596, center, 4C31.04, and at right, 1946+708.  All error bars
plotted here are $1 \sigma$.  The dotted line in the top panels represents the
best fit ``nuker'' model (see Table 2 and \S 3.2 for details). In the cases of
4C 31.04 and 1946+708 we have excluded  regions where the parameter fits did
not converge or gave very large errors.}

\end{figure}

\vfill\eject

\begin{figure}

\centerline{\includegraphics*[scale=0.7]{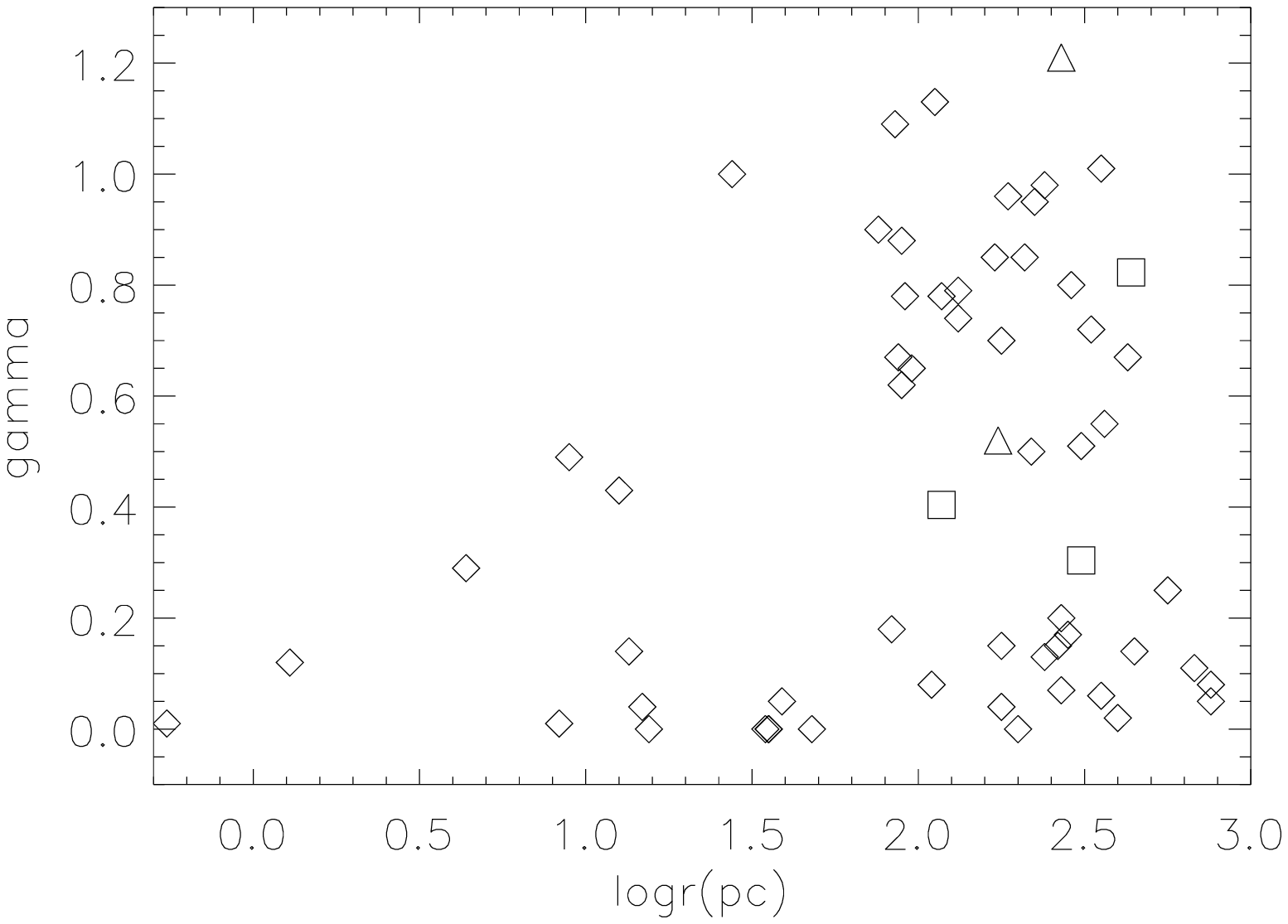}}
\centerline{\includegraphics*[scale=0.7]{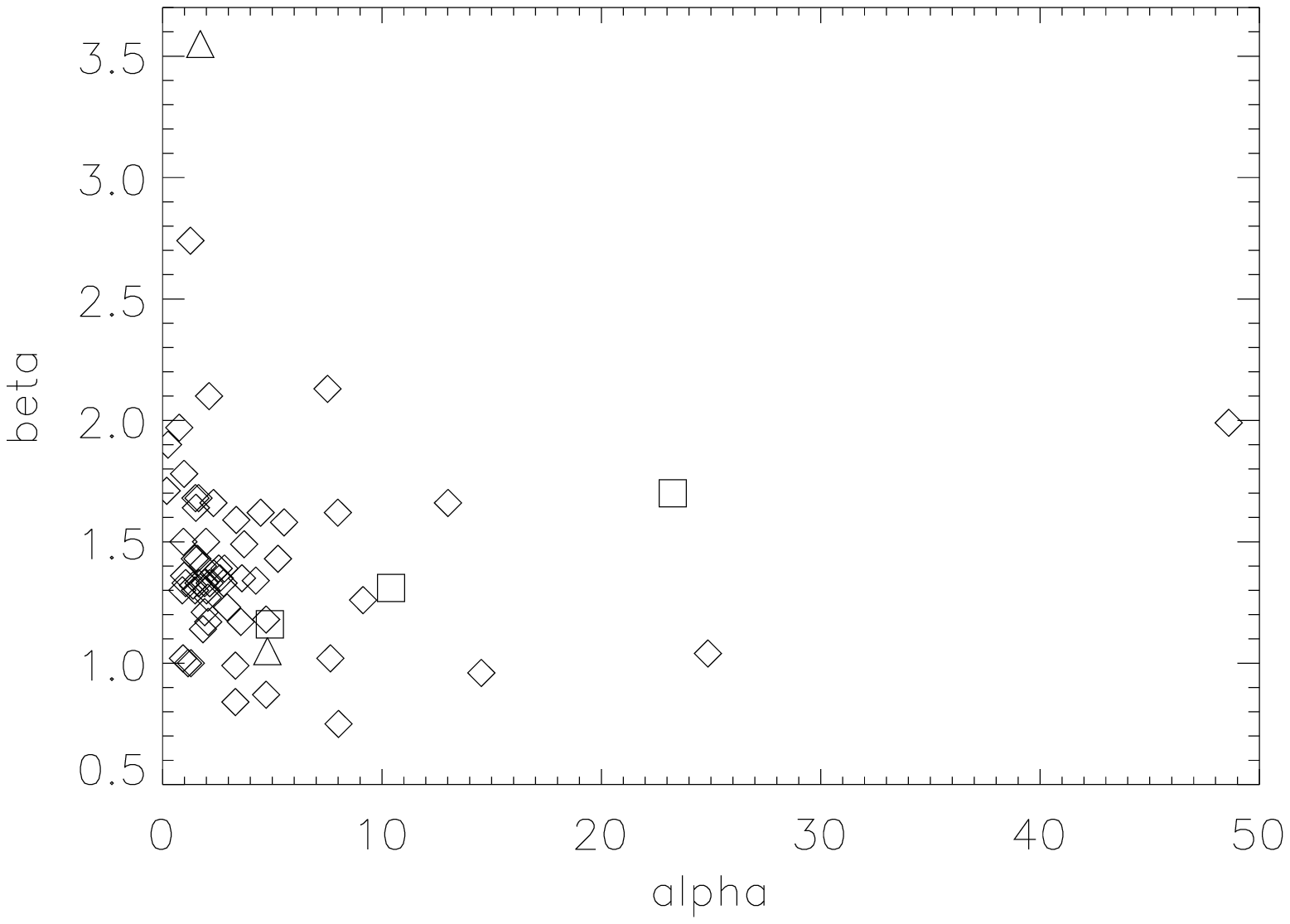}}

\caption[]{Nuker fit parameters for these three objects, compared to the
objects in the Faber et al. (1997) sample.  Our objects are the squares in this
plots, and the Faber et al. objects are  the diamonds, while the triangles are
Faber et al.'s simulations of M31 and  M32 (respectively) at the distance of
the Virgo cluster. While these objects appear to be rather intermediate between
the ``core'' and ``power-law'' galaxy types, this is likely a product of their
several times greater distance compared to  the objects in the Faber et al.
sample.  See \S 4 for discussion.}

\end{figure}

\vfill\eject

\begin{figure}

\centerline{\includegraphics*[scale=1.00]{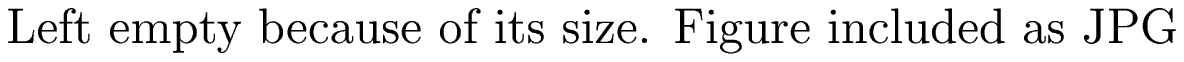}}

\caption[]{Total Extinction $(A_B)$ Images.  At left, 1146+596 (greyscale  runs
from 0 to 1 mag, corresponding to $N_H$ values ranging from 0 to 1.4$\times
10^{21} {\rm cm^{-2}}$) ; at middle, 4C31.04 (greyscale runs from 0 to 2 mag,
corresponding to $N_H$ values ranging from 0 to 2.8 $\times 10^{21} {\rm
cm^{-2}}$); and at right, 1946+708 (greyscale runs from 0 to 2 mag,
corresponding to $N_H$ values ranging from 0 to 2.8 $\times 10^{21} {\rm
cm^{-2}}$).  North is at the  top and east is at the left in all three images.}



\end{figure}

\end{document}